\documentclass[10pt,aps,preprintnumbers,prc,noshowpacs,nofootinbib,noshowkeys,floatfix,superscriptaddress]{revtex4-2}
\usepackage{graphicx}
\usepackage{cancel}
\usepackage[colorlinks=true,linktocpage=true,linkcolor=blue,citecolor=blue]{hyperref}
\usepackage[usenames,dvipsnames]{color}
\usepackage{combelow}
\usepackage{amsmath, amssymb,oldgerm}
\usepackage{mathtools}
\usepackage{multirow}
\usepackage{longtable}
\usepackage{xcolor}
\usepackage{dsfont}
\usepackage[normalem]{ulem}  
\usepackage{braket}
\usepackage{slashed}
\usepackage{hyperref}
\usepackage[mathscr]{eucal}
\usepackage{tikz}
\usepackage{pgfplots}
\usepackage[compat=1.1.0]{tikz-feynman}

\newcommand{\s}{y}

\newcommand{\mC}{\mathfrak{C}}

\newcommand{\beq}{\begin{equation}}
\newcommand{\eeq}{\end{equation}}

\newcommand{\eqs}{Eqs.~}
\newcommand{\eq}{Eq.~}

\renewcommand{\k}{\mathbf{k}}

\renewcommand{\d}{\mathrm{d}}
\renewcommand{\s}{\mathfrak{s}}

\newcommand{\R}{\mathrm{Re}}
\newcommand{\avg}[1]{\left\langle#1\right\rangle}

\begin{document}

\title{Generating Tensor Polarization from Shear Stress}

\author{David Wagner}
\affiliation{Institute for Theoretical Physics, Goethe University,
Max-von-Laue-Str.\ 1, D-60438 Frankfurt am Main, Germany} 
\affiliation{Department of Physics, West University of Timi\cb{s}oara, \\
Bd.~Vasile P\^arvan 4, Timi\cb{s}oara 300223, Romania}
 
\author{Nora Weickgenannt}
\affiliation{Institute for Theoretical Physics, Goethe University,
Max-von-Laue-Str.\ 1, D-60438 Frankfurt am Main, Germany} 
\affiliation{Institut de Physique Théorique, Université Paris Saclay,\\
CEA, CNRS, F-91191 Gif-sur-Yvette}

\author{Enrico Speranza}
\affiliation{Illinois Center for Advanced Studies of the Universe and Department of Physics, University of Illinois at Urbana-Champaign, Urbana, Illinois 61801, USA}

\begin{abstract}
We derive an expression for the tensor polarization of a system of massive spin-1 particles in a hydrodynamic framework. Starting from quantum kinetic theory based on the Wigner-function formalism, we employ a modified method of moments which also takes into account all spin degrees of freedom. It is shown that the tensor polarization of an uncharged fluid is determined by the shear-stress tensor. In order to quantify this novel polarization effect, we provide a formula which can be used for numerical calculations of vector-meson spin alignment in relativistic heavy-ion collisions.
\end{abstract}
\maketitle

\section{Introduction}
The observation of polarization phenomena in relativistic heavy-ion collisions has opened a new direction of research in the physics of the hot and dense nuclear matter \cite{Becattini:2020ngo,Becattini:2022zvf}. The STAR Collaboration showed that $\Lambda$-baryons emitted in noncentral nuclear collisions are spin polarized along the direction of the global angular momentum~\cite{STAR:2017ckg,Adam:2018ivw}. This finding provides the evidence that in the quark-gluon plasma particle spin polarization is triggered by rotation [as suggested in Refs.~\cite{Liang:2004ph,Voloshin:2004ha,Betz:2007kg,Becattini:2007sr}] in a way which resembles the time-honored Barnett effect \cite{Barnett:1935}. Despite early success in describing global polarization data \cite{Becattini:2007sr,Becattini:2013vja,Becattini:2013fla,Becattini:2015ska,Becattini:2016gvu,Karpenko:2016jyx,Pang:2016igs,Xie:2017upb}, discrepancies between theory and experiment triggered big theoretical efforts both at the phenomenological \cite{Becattini:2017gcx,Florkowski:2019qdp,Florkowski:2019voj,Zhang:2019xya,Becattini:2019ntv,Xia:2019fjf,Wu:2019eyi,Sun:2018bjl,Liu:2019krs,Florkowski:2021wvk,Liu:2021uhn,Fu:2021pok,Becattini:2021suc,Becattini:2021iol} and more formal level with the formulation of relativistic spin hydrodynamics \cite{Florkowski:2017ruc,Florkowski:2017dyn,Florkowski:2018myy,Florkowski:2018fap,Weickgenannt:2019dks,Bhadury:2020puc,Weickgenannt:2020aaf,Shi:2020htn,Speranza:2020ilk,Bhadury:2020cop,Singh:2020rht,Bhadury:2021oat,Peng:2021ago,Sheng:2021kfc,Sheng:2022ssd,Hu:2021pwh,Hu:2022lpi,Singh:2022ltu,Montenegro:2017rbu,Montenegro:2018bcf,Montenegro:2020paq,Gallegos:2021bzp,Hattori:2019lfp,Fukushima:2020ucl,Li:2020eon,She:2021lhe,Wang:2021ngp,Wang:2021wqq,Daher:2022xon,Gallegos:2020otk,Garbiso:2020puw,Cartwright:2021qpp,Hongo:2021ona,Weickgenannt:2022zxs,Weickgenannt:2022jes,Gallegos:2022jow,Bhadury:2022qxd,Cao:2022aku}.
More recently, experimental studies of the so-called spin alignment of massive spin-1 particles such as $\phi$ and $K^{\star 0}$ mesons have been also carried out \cite{ALICE:2019aid,Mohanty:2021vbt,STAR:2022fan}. The data shows that the spin alignment is much larger compared to theoretical predictions given by models based on the assumption of local equilibrium \cite{Becattini:2007sr}. This poses a new puzzle which is currently the subject of intense work \cite{Liang:2004xn,Yang:2017sdk,Sheng:2019kmk,Sheng:2020ghv,Xia:2020tyd,Goncalves:2021ziy,Muller:2021hpe,Sheng:2022wsy,Sheng:2022ffb} for which, however, an established solution is still missing. 

In heavy-ion experiments, the spin vector polarization of $\Lambda$-baryons can be directly extracted from the angular distribution of their weak decay \cite{STAR:2017ckg,Adam:2018ivw}. The case of massive spin-1 particles is different. First, it is important to note that the polarization state of a vector meson is fully specified by three parameters corresponding to the conventional vector polarization and by additional five parameters called tensor polarization \cite{Leader:2001}. In fact, tensor polarization is a property which characterizes only particles with spin higher than 1/2. In general, vector and tensor polarization are independent quantities and, therefore, one can have a spin-1 particle which is tensor polarized and not vector polarized, and vice versa \cite{Leader:2001}. In experiments, since for vector mesons only parity-conserving decays are studied \cite{ALICE:2019aid,Mohanty:2021vbt}, the spin alignment only gives information on the tensor polarization state. 

In Refs. \cite{Baym:2017qxy,Speranza:2018osi} it was shown that vector mesons emitted from a thermalized medium are in general tensor polarized even if the system is in global equilibrium without rotation. Such tensor polarization is due to the imbalance between transverse and longitudinal spectral functions \cite{Baym:2017qxy,Speranza:2018osi}. In this paper, we propose a different mechanism. We consider an uncharged fluid composed of massive spin-1 particles near local thermodynamic  equilibrium. In our framework, tensor polarization arises due the presence of shear stress in the fluid. An intuitive explanation can be given only based on parity arguments. Since tensor polarization is a parity-even rank-2 traceless and symmetric tensor \cite{Leader:2001}, in a hydrodynamic framework it can only be proportional to the shear stress tensor of the fluid at first order in deviations from equilibrium. In this work we derive the expression for the tensor polarization starting from quantum kinetic theory for massive spin-1 particles. In order to calculate the dissipative corrections, we use the method of moments. In particular, we define new rank-2 spin moments which extend the previous formulations for the spin-0 \cite{Denicol:2012cn} and spin-1/2 cases \cite{Weickgenannt:2022zxs}. 

Our notation and conventions are: $a\cdot b\coloneqq a^\mu b_\mu$,
$a_{[\mu}b_{\nu]}\coloneqq a_\mu b_\nu-a_\nu b_\mu$, $a_{(\mu}b_{\nu)}\coloneqq a_\mu b_\nu+a_\nu b_\mu$, $g_{\mu \nu} \coloneqq \mathrm{diag}(+,-,-,-)$,
$\epsilon^{0123} = - \epsilon_{0123} \coloneqq 1$. The $\ell$-th rank projector onto the subspace of traceless symmetric tensors orthogonal to the fluid 4-velocity $u^\mu$~\cite{DeGroot:1980dk} is denoted as $\Delta^{\mu_1\cdots\mu_\ell}_{\nu_1\cdots\nu_\ell}$, and we write a projected tensor $A$ as $A^{\langle\mu_1 \cdots \mu_\ell \rangle} \coloneqq \Delta^{\mu_1\cdots \mu_\ell}_{\nu_1\cdots \nu_\ell} A^{\nu_1\cdots\nu_\ell}$.

\section{Kinetic theory for vector particles}

Let us consider the Lagrangian for a Proca field $V^\mu$ of mass $m$,
\begin{equation}
\label{Lagrangian}
\mathcal{L}=-\hbar  \left(\frac12 V^{\dagger\mu\nu} V_{\mu\nu} -\frac{m^2}{\hbar^2}V^{\dagger\mu}V_\mu\right)+\mathcal{L}_{\text{int}}\;.
\end{equation}
where $\mathcal{L}_{\text{int}}$ is a general interaction Lagrangian.
The fundamental object of quantum kinetic theory is the Wigner function defined as \cite{Vasak:1987um, Elze:1986hq,Elze:1989un,Huang:2020kik,Hattori:2020gqh,Weickgenannt:2022jes}
\begin{equation}
\label{Wigner_function}
W^{\mu\nu}(x,k)\coloneqq-\frac{2}{(2\pi\hbar)^4\hbar} \int \d^4 y\, e^{-ik\cdot y/\hbar} \avg{:V^{\dagger\mu}\left(x+y/2\right) V^\nu\left(x-y/2\right):}\;,
\end{equation}
where $\avg{:\cdots:}$ denotes the normal-ordered ensemble average. This Wigner-transform of the two-point function defines a quantum analogue of the distribution function known from classical kinetic theory. Assuming that quantum effects are small (meaning that the Compton wavelength of the particles has to be small compared to a typical macroscopic length scale), one can perform a so-called $\hbar$-expansion, i.e., write
\begin{equation}
    W^{\mu\nu}(x,k)=W^{(0),\mu\nu}(x,k)+\hbar W^{(1),\mu\nu}(x,k)+\cdots\;,
\end{equation}
where the Planck constant acts as a book-keeping parameter. In the following, all results are derived from employing such an expansion up to first order in $\hbar$. 
Note that in Eq. \eqref{Wigner_function} the momentum variable $k$ is not necessarily on the mass shell. However, one can show \cite{Weickgenannt:2021cuo, Sheng:2021kfc, forth44} that, to first order in the $\hbar$-expansion, the off-shell terms cancel in the evolution equation of the Wigner function, such that it is sufficient to consider the part that is on shell. Considering the fact that a charged vector field has 3 (complex) independent components \cite{Weinberg:1995mt}, it is evident that the Wigner function must have 9 independent degrees of freedom, while the remaining 7 components can be expressed in terms of these \cite{Weickgenannt:2022jes}. These degrees of freedom can be shown to consist of a scalar (1 component), a pseudovector (3 components), and a traceless symmetric tensor (5 components). As shown in Appendix  \ref{app:Wigner_to_obs}, the pseudovector degree of freedom can be related to the vector polarization of the particles, while the traceless symmetric tensor corresponds to the tensor polarization. A convenient way to treat these 9 independent components in a compact fashion is to enlarge the phase-space by introducing an additional ``spin'' variable $\s^\mu$ \cite{Weickgenannt:2020aaf}, together with a respective measure
\begin{equation}
\d S(k)\coloneqq \frac{3m}{2\sigma \pi}\d^4 \s \,\delta(\s^2+\sigma^2)\delta(k\cdot \s)\;, \quad\sigma^2 \coloneqq2\;.
\end{equation}
Note that we have the following identities,
\begin{equation}
    \int \d S(k) =3\;,\quad \int \d S(k) \s^\mu \s^\nu =-2K^{\mu\nu}\;,\quad \int \d S(k) K^{\mu\nu}_{\alpha\beta} \s^\alpha \s^\beta \s_\rho \s_\sigma =\frac85 K^{\mu\nu}_{\rho\sigma}\;,
\end{equation}
while the integral over any odd number of spin vectors vanishes.
Here, $K^{\mu\nu}\coloneqq g^{\mu\nu}-k^\mu k^\nu /m^2$ and $K^{\mu\nu}_{\rho\sigma}\coloneqq 1/2K^{\mu}_{(\rho}K^\nu_{\sigma)}-1/3 K^{\mu\nu}K_{\rho\sigma}$ denote the projectors onto subspaces irreducible with respect to the little group of $k^\mu$.
We can then define a scalar distribution function \cite{Weickgenannt:2022jes}
\begin{equation}
f(x,k,\s)\coloneqq  H^{\nu\mu}(k,\s) W^{\mathrm{on-shell}}_{\mu\nu}(x,k)\;, \quad H^{\mu\nu}(k,\s)\coloneqq \frac13 K^{\mu\nu}+\frac{i}{2}\epsilon^{\mu\nu\alpha\beta}\frac{k_\alpha}{m}\s_\beta +\frac58 K^{\mu\nu}_{\alpha\beta} \s^{\alpha} \s^{\beta}\;,
\label{eq:def_f}
\end{equation}
where $W^{\mathrm{on-shell}}_{\mu\nu}$ denotes the part of the Wigner function proportional to $\delta(k^2-m^2)$. 
It is important to note that, to first order in the $\hbar$-expansion, the distribution function $f(x,k,\s)$ contains the complete information necessary to reconstruct the full Wigner function. In the noninteracting case, the following inverse relation also holds,
\begin{equation}
    W_{\mu\nu}^{\text{on-shell}}(x,k)=\int \d S(k) h_{\mu\nu}(k,\s) f(x,k,\s)\;,\quad h_{\mu\nu}(k,\s)\coloneqq \frac13 K_{\mu\nu}+\frac{i}{2}\epsilon_{\mu\nu\alpha\beta}\frac{k^\alpha}{m}\s^\beta + K_{\mu\nu}^{\alpha\beta} \s_{\alpha} \s_{\beta}\;.
\end{equation}
Starting from the equations of motion for the vector field that follow from the Lagrangian \eqref{Lagrangian}, it can be shown that the evolution equation of the phase-space distribution function reads \cite{forth44}
\begin{equation}
\label{eq_bol}
k\cdot \partial f(x,k,\s)=\mathfrak{C}[f] \; ,
\end{equation}
where
\begin{align}
\mathfrak{C}[f] &\coloneqq \frac12 \int \d \Gamma_1 \, \d \Gamma_2 \, \d \Gamma' \,\d \bar{S}(k) (2\pi\hbar)^4 \delta^{(4)}(k+k'-k_1-k_2) \mathcal{W}\nonumber\\
&\times\left[f(x+\Delta_1-\Delta,k_1,\s_1)f(x+\Delta_2-\Delta,k_2,\s_2)-f(x+\Delta'-\Delta,k',\s')f(x,k,\bar{\s})\right] 
\label{C_final}
\end{align}  
and we introduced the $(x,k,\s)$-phase-space measures 
\begin{equation}
    \d \Gamma \coloneqq \d K \d S(k)\;,\quad \d K \coloneqq  \frac{2}{(2\pi\hbar)^3}\d^4 k \,\delta(k^2-m^2)\;.
\end{equation}
The transition rate is given by
\begin{eqnarray}
\mathcal{W}\coloneqq  \frac{(2\pi\hbar)^3}{32} M^{\gamma_1\gamma_2\delta_1\delta_2}M^{\zeta_1\zeta_2\eta_1\eta_2} h_{1,\gamma_1\eta_1} h_{2,\gamma_2\eta_2} h'_{\zeta_2\delta_2} \left(H_{\zeta_1}{}^{\alpha} \bar{h}_{\alpha\delta_1}+\bar{h}_{\zeta_1}{}^{\alpha} H_{\alpha\delta_1}\right)\;,\label{W_ps}
\end{eqnarray}
while the vectors $\Delta_1$, $\Delta_2$, $\Delta'$ and $\Delta$ read 
\begin{subequations}\label{eq:def_Delta}
\begin{eqnarray}
    \Delta_1^\mu&\coloneqq& \frac23\frac{1}{\mathcal{W}}\frac{(2\pi\hbar)^3}{64}\frac{i\hbar}{2m^2} M^{\gamma_1\gamma_2\delta_1\delta_2}M^{\zeta_1\zeta_2\eta_1\eta_2} \left(h^\mu_{1}{}_{\eta_1}k_{1,\gamma_1}-k_{1,\eta_1}h_{1,\gamma_1}{}^\mu\right)h_{2,\gamma_2\eta_2}h'_{\zeta_2\delta_2}H_{\zeta_1\delta_1} \;,\\
    \Delta_2^\mu&\coloneqq& \frac23 \frac{1}{\mathcal{W}}\frac{(2\pi\hbar)^3}{64}\frac{i\hbar}{2m^2} M^{\gamma_1\gamma_2\delta_1\delta_2}M^{\zeta_1\zeta_2\eta_1\eta_2} h_{1,\gamma_1\eta_1}\left(h^\mu_{2}{}_{\eta_2}k_{2,\gamma_2}-k_{2,\eta_2}h_{2,\gamma_2}{}^\mu\right)h'_{\zeta_2\delta_2}H_{\zeta_1\delta_1} \;,\\
    \Delta'^\mu&\coloneqq& \frac23\frac{1}{\mathcal{W}}\frac{(2\pi\hbar)^3}{64}\frac{i\hbar}{2m^2} M^{\gamma_1\gamma_2\delta_1\delta_2}M^{\zeta_1\zeta_2\eta_1\eta_2} h_{1,\gamma_1\eta_1}h_{2,\gamma_2\eta_2}\left(h'^\mu{}_{\delta_2}k'_{\zeta_2}-k'_{\delta_2}h'_{\zeta_2}{}^\mu\right)H_{\zeta_1\delta_1} \;,\\
    \Delta^\mu&\coloneqq& \frac23\frac{1}{\mathcal{W}}\frac{(2\pi\hbar)^3}{64}\frac{i\hbar}{2m^2} M^{\gamma_1\gamma_2\delta_1\delta_2}M^{\zeta_1\zeta_2\eta_1\eta_2} h_{1,\gamma_1\eta_1}h_{2,\gamma_2\eta_2}h'_{\zeta_2\delta_2}\left(H^\mu{}_{\delta_1}k_{\zeta_1}-k_{\delta_1}H_{\zeta_1}{}^\mu\right) \;,
\end{eqnarray}
\end{subequations}
where we abbreviated $h_1\coloneqq h(k_1,\s_1)$ (and analogously for $h_2$, $h'$, $\bar{h}$ and $H$). The vectors \eqref{eq:def_Delta} denote shifts in the particle position from the point $x$, characterizing the nonlocality of the collision. It has been shown in Refs. \cite{Weickgenannt:2020aaf,Weickgenannt:2021cuo} that this nonlocality is essential to explain the spin-polarization of particles, as it introduces a nonvanishing orbital angular momentum into the collision that can then be converted into spin, since the total angular momentum is conserved. However, it will become clear in Sec. \ref{sec:tens_pol_from_pi} that the tensor polarization of vector particles does not depend on these nonlocalities, but arises from purely local effects.
In Eqs. \eqref{W_ps} and \eqref{eq:def_Delta}, $M$ is the tree-level vertex of the theory, and is related to the transfer matrix elements via
\begin{equation}
\bra{k,k';\lambda,\lambda'} \hat{t}\ket{k_1,k_2;\lambda_1,\lambda_2}
=  \epsilon_\mu^{*(\lambda)}(k)\epsilon_\nu^{*(\lambda')}(k') \epsilon_\alpha^{(\lambda_1)}(k_1) \epsilon_\beta^{(\lambda_2)}(k_2) M^{\mu\nu\alpha\beta}\;,
    \label{eq:rel_t_M}
\end{equation}
where e.g. $\ket{k_1,k_2;\lambda_1,\lambda_2}$ denotes a two-particle state with momenta $(k_1,k_2)$ and spins $(\lambda_1,\lambda_2)$, while $\epsilon^{(\lambda)}_\mu (k)$ is the polarization vector of a vector particle with momentum $k$ and spin $\lambda$.
Note that the form of the Boltzmann equation~\eqref{eq_bol} and \eqref{C_final} for binary elastic collisions closely resembles the formulation presented in Refs. \cite{Weickgenannt:2020aaf, Weickgenannt:2021cuo, Wagner:2022amr}.

\section{Relativistic Hydrodynamics and tensor polarization}

We consider an uncharged fluid with spin degrees of freedom and tensor polarization governed by  the conservation equations 
\begin{equation}
\partial_\mu T^{\mu\nu}=0\;,\quad \hbar\partial_\lambda S^{\lambda,\mu\nu}=T^{[\nu\mu]}\;,
\end{equation}
where $T^{\mu\nu}$ is the energy-momentum tensor and $S^{\lambda,\mu\nu}$ is the spin tensor. 
In this work we choose the Hilgevoord-Wouthuysen (HW) pseudo-gauge up to first order in $\hbar$ \cite{Speranza:2020ilk,Weickgenannt:2022jes}, 
\begin{equation}
    T^{\mu\nu}\coloneqq \int \d\Gamma\, k^\mu k^\nu f(x,k,\s)\; , \qquad S^{\lambda,\mu\nu}\coloneqq \int \d\Gamma\, k^\lambda \left(\Sigma_\s^{\mu\nu}-\frac{\hbar}{3m^2}k^{[\mu}\partial^{\nu]}\right)f(x,k,\s) \; ,\label{consquan}
\end{equation}
where we defined $\Sigma^{\mu\nu}_\s \coloneqq -(1/m)\epsilon^{\mu\nu\alpha\beta}k_\alpha \s_\beta$.
The (momentum-dependent) tensor polarization is given by
\begin{equation}
  \Theta^{\mu\nu}(k)= {\frac12 \sqrt{\frac32}}\frac{1}{N(k)}  \int \d \Sigma_\lambda k^\lambda \int \d S(k) K^{\mu\nu}_{\alpha\beta} \s^\alpha\s^\beta f(x,k,\s) \;,\label{tensor_pol0}
\end{equation}
where the prefactor is defined in accordance with Ref.~\cite{Leader:2001}, $N(k)\coloneqq  \int \d\Sigma_\lambda k^\lambda K_{\alpha\beta} W^{\alpha\beta} $ and $\d \Sigma_\lambda$ denotes integration over a spacelike hypersurface, which, for example, can be taken to be the freeze-out hypersurface. As will be shown later, this quantity is related to the spin alignment measured in experiments \cite{ALICE:2019aid,Mohanty:2021vbt,STAR:2022fan}. A derivation of Eq.\ \eqref{tensor_pol0} is provided in Appendix \ref{app:Wigner_to_obs}.

\subsection{Moment expansion}

In order to determine the dissipative corrections to the tensor polarization, we extend the formalism developed in Ref.\ \cite{Denicol:2012cn} for spin-0 particles and in Ref.\ \cite{Weickgenannt:2022zxs} for spin-1/2 particles to the case of spin 1. We split the distribution function $f(x,k,\s)$ into a local-equilibrium and a dissipative contribution,
\begin{equation}
    f(x,k,\s)=f_\text{eq}(x,k,\s)+\delta f_{\k\s} \label{splitf}
\end{equation}
with the local-equilibrium part \cite{Weickgenannt:2021cuo}
\begin{equation}
f_\text{eq}(x,k,\s)\coloneqq \exp\left(-\beta_0 E_\k -\frac{\hbar}{2m}\epsilon^{\mu\nu\alpha\beta}\Omega_{\mu\nu} k_\alpha \s_\beta   \right)\;,
\label{f_eq}
\end{equation}
 where $E_\k\coloneqq k\cdot u$. Note again that \eq\eqref{f_eq} as well as all calculations in this paper are valid up to first order in $\hbar$. The Lagrange multipliers for the four-momentum and total angular momentum are given by $\beta_0 u^\mu$ and  $\Omega_{\mu\nu}$, respectively, with $\beta_0$ being the inverse temperature, $u^\mu$ the fluid four-velocity and $\Omega_{\mu\nu}$ the spin potential. Since the tensor polarization is not related to any conserved quantity, it does not appear in the local-equilibrium distribution function.
The deviation from local equilibrium $\delta f_{\k\s}$ is first expanded in the spin variable $\s^\mu$, where it is at most bilinear, cf. Eq. \eqref{eq:def_f}. Thus we can write
\begin{equation}
    \delta f_{\k \s} = f_{0\k} \left(\phi_\k -\s_\mu \zeta_\k^\mu + \s_\alpha \s_\beta K^{\alpha\beta}_{\mu\nu} \xi_\k^{\mu\nu} \right)\;,
    \label{eq:delta_f_expansion_1}
\end{equation}
where $f_{0\k}\coloneqq\exp(-\beta_0 E_\k)$ is the zeroth-order equilibrium distribution function.
Here we assumed $\zeta_\k^\mu$ and $\xi_\k^{\mu\nu}$ to be orthogonal to the four-momentum and (in the case of $\xi_\k^{\mu\nu}$) traceless, which can be done without loss of generality due to the symmetries of $\s^\mu$ and $K^{\mu\nu}_{\alpha\beta} \s^\alpha\s^\beta$ \cite{Weickgenannt:2022zxs}.
Then, it is possible to explicitly use these properties to eliminate the components of $\zeta_\k^\mu$ and $\xi_\k^{\mu\nu}$ that are parallel to the fluid four-velocity $u^\mu$, obtaining
\begin{equation}
\delta f_{\k \s}=f_{0\k} \left(\phi_\k -\s^\nu \Xi_{\nu\mu} \zeta_\k^\mu + \s_\alpha \s_\beta K^{\alpha\beta}_{\mu\nu} \Xi^{\mu\nu}_{\rho\sigma} \xi_\k^{\rho\sigma} \right)\;,
\label{eq:delta_f_expansion_2}
\end{equation}
where we defined the tensors 
\begin{equation}
\Xi_{\mu\nu}\coloneqq \Delta_{\mu\nu}+\frac{k_{\langle\mu\rangle} k_{\langle\nu\rangle}}{E_\k^2}\;,\quad \Xi_{\mu\nu,\alpha\beta}\coloneqq \frac12 \left(\Xi_{\mu\alpha}\Xi_{\nu\beta}+\Xi_{\mu\beta}\Xi_{\nu\alpha}\right)-\frac{1}{\Xi^2}\Xi_{\mu\gamma}\Xi^{\;\;\gamma}_{\nu}\Xi_{\delta\alpha}\Xi^{\delta}_{\;\;\beta}
\label{eq:def_Xi}
\end{equation}
with $\Xi^2\coloneqq \Xi^{\mu\nu}\Xi_{\mu\nu}=2+m^4/E_\k^4$.
Expanding $\phi_\k$, $\zeta_\k^\mu$ and $\xi_\k^{\mu\nu}$ terms of irreducible moments, we find
\begin{align}
\delta f_{\k\s}&=f_{0\k} \sum_{\ell=0}^\infty  k_{\langle\mu_1}\cdots k_{\mu_\ell\rangle} \left(\sum_{n\in \mathbb{S}_\ell^{(0)}} \mathcal{H}^{(0,\ell)}_{\k n}\rho_n^{\mu_1\cdots\mu_\ell}\right.\nonumber\\
&-\left.\s^\nu \Xi_{\nu\mu} \sum_{n\in \mathbb{S}_\ell^{(1)}} \mathcal{H}^{(1,\ell)}_{\k n}\tau_n^{\langle\mu\rangle,\mu_1\cdots \mu_\ell}+ \s_\alpha \s_\beta  K^{\alpha\beta}_{\mu\nu}\Xi^{\mu\nu}_{\rho\sigma} \sum_{n\in \mathbb{S}_\ell^{(2)}} \mathcal{H}^{(2,\ell)}_{\k n}\psi_n^{\langle\rho\sigma\rangle,\mu_1\cdots \mu_\ell}  \right)\;.\label{delta_f_expansion}
\end{align} 
Here $\mathbb{S}_\ell^{(n)}$ denotes the set of moments of tensor-rank $\ell$ in momentum and $n$ in spin that are included in the theory, and the irreducible moments are given by
\begin{subequations}\label{eq:mom_def_all}
\begin{align}
\rho_r^{\mu_1\cdots\mu_\ell}&\coloneqq\int \d \Gamma E_\k^r k^{\langle\mu_1} \cdots k^{\mu_\ell \rangle} \delta f_{\k\s}\;,\\
\tau_r^{\mu, \mu_1\cdots\mu_\ell}&\coloneqq\int \d \Gamma E_\k^r\s^\mu k^{\langle\mu_1} \cdots k^{\mu_\ell \rangle} \delta f_{\k\s}\;,\label{taudef}\\
\psi_r^{\mu\nu,\mu_1\cdots\mu_\ell}&\coloneqq\int \d \Gamma E_\k^r K^{\mu\nu}_{\alpha\beta} \s^\alpha \s^\beta   k^{\langle\mu_1} \cdots k^{\mu_\ell \rangle} \delta f_{\k\s}\; . \label{psidef}
\end{align} 
\end{subequations}
Note that, as was the case in Ref. \cite{Weickgenannt:2022zxs}, due to the explicit removal of redundant degrees of freedom [cf. Eq. \eqref{eq:delta_f_expansion_2}] only moments orthogonal to the four-velocity in all indices enter the expansion \eqref{delta_f_expansion}.
Furthermore, we introduced the polynomials
\begin{equation}
    \mathcal{H}^{(j,\ell)}_{\k n}\coloneqq \frac{(2j+1)!!}{2^j j!} \frac{W^{(\ell)}}{\ell!} \sum_{m\in \mathbb{S}_{\ell}^{(j)}} \sum_{q=0}^m a^{(\ell)}_{mn}a^{(\ell)}_{mq}E_\k^q\;,
\end{equation}
where the coefficients $a_{mn}^{(\ell)}$ are constructed via Gram-Schmidt orthogonalization, cf. Ref. \cite{Denicol:2012cn}. The normalization reads $W^{(\ell)}\coloneqq(-1)^\ell / I_{2\ell,\ell}$, where we defined the standard thermodynamic integrals
\begin{equation}
    I_{nq}\coloneqq \frac{1}{(2q+1)!!} \int \d \Gamma E_\k^{n-2q} \left(E_\k^2 -m^2\right)^{q}f_{0\k}\;.
\end{equation}
The rank-(2+$\ell$) tensors in \eq\eqref{psidef} are new compared to the previously developed hydrodynamic framework for spin-0 and spin-1/2 particles \cite{Denicol:2012cn,Weickgenannt:2022zxs} and correspond to dissipative degrees of freedom associated with tensor polarization, cf. Sec. \ref{sec:tens_pol_from_pi}. Inserting \eq\eqref{splitf} into \eq\eqref{eq_bol}, the Boltzmann equation takes the form
\begin{equation}
\delta \dot{f}_{\k\s} + \dot{f}_{0\k}+E_\k^{-1} k\cdot \nabla f_{0\k}+E_\k^{-1} k\cdot \nabla \delta f_{\k\s} =E_\k^{-1} \mathfrak{C}[f]\;, \label{bolbi}
\end{equation}
which is the starting point for the derivation of the equations of motion for the irreducible moments. For the purpose of this paper
the full set of coupled equations of motion is not needed and we will only focus on the tensor-polarization moments $\psi_r^{\mu\nu,\mu_1\cdots\mu_\ell}$. Integrating \eq\eqref{bolbi} over $\int \d S(k) K^{\mu\nu}_{\alpha\beta}\s^\alpha\s^\beta$, we obtain equations of motion of the form 
\begin{eqnarray}
\dot{\psi}_r^{\langle\mu\nu\rangle,\langle\mu_1\cdots \mu_\ell\rangle} -C_{r-1}^{\langle\mu\nu\rangle,\langle\mu_1\cdots\mu_\ell\rangle}&=&\mathcal{O}( \R^{-1}\partial)^{\langle\mu\nu\rangle,\mu_1\cdots\mu_\ell}\;.\label{relax_psi}
\end{eqnarray}
Here we used that $\int \d S(k) K^{\mu\nu}_{\alpha\beta}\s^\alpha\s^\beta f_\text{eq}(x,k,\s)=0$, which follows from \eq\eqref{f_eq}. This implies that tensor polarization vanishes in equilibrium up to first order in $\hbar$. The contributions from the last term on the left-hand side of \eq\eqref{bolbi} to \eq\eqref{relax_psi}, denoted by $\mathcal{O}( \R^{-1}\partial)$, correspond to quantities linear in gradients of dissipative quantities, i.e., of first order in  the so-called inverse Reynolds numbers $\R^{-1}$. Note that the first term on the left-hand side of \eq\eqref{relax_psi} is also of order $\mathcal{O}( \R^{-1}\partial)$.
Furthermore, we defined the generalized collision integrals
\begin{eqnarray}
C_r^{\mu\nu,\langle \mu_1 \cdots \mu_\ell\rangle}&\coloneqq& \int \d \Gamma E_\k^r K^{\mu\nu}_{\alpha\beta} \s^\alpha \s^\beta k^{\langle\mu_1}\cdots k^{\mu_\ell\rangle} \mC[f]\;.\label{irred_coll}
\end{eqnarray}
The explicit form of the right-hand side of \eq\eqref{relax_psi} is of no importance for the following discussion, since we will assume that the tensor-polarization moments are given by their Navier-Stokes values, which are determined by neglecting contributions of order $\mathcal{O}( \R^{-1}\partial)$ in \eq\eqref{relax_psi}. This is justified since, in contrast to the components of the energy-momentum tensor or spin tensor, the tensor-polarization moments are not part of the conserved quantities \eqref{consquan}. Therefore, it is not necessary to treat them dynamically in second-order hydrodynamics, and it is reasonable to expect that the Navier-Stokes values will constitute the leading-order contribution, while possible second-order terms would lead to small corrections.

\subsection{Truncation}
\label{subsec:truncation}

Since we expect the conserved quantities \eqref{consquan} to dominate the evolution of the system on long time scales, it is reasonable to take the irreducible moments appearing there as the dynamical degrees of freedom of our theory. Decomposing the energy-momentum tensor with respect to the fluid velocity $u^\mu$ as 
\begin{equation}
T^{\mu\nu}= \epsilon u^\mu u^\nu -\Delta^{\mu\nu}(P_0+\Pi)+\pi^{\mu\nu}\; ,
\end{equation}
where $\epsilon$ is the energy-density, $P_0$ is the isotropic pressure, $\Pi$ is the bulk-viscous pressure, $\pi^{\mu\nu}$ denotes the shear-stress tensor, and imposing the Landau frame condition $ T^{\mu\nu}u_\nu=\epsilon u^\mu$ as well as the matching condition $u_\mu u_\nu T^{\mu\nu}=u_\mu u_\nu T^{\mu\nu}_\text{eq}$, we identify
the dynamical moments $\rho_0\equiv-(3/m^2)\Pi$ and $\rho_0^{\mu\nu}\equiv\pi^{\mu\nu}$, while $\rho_1=\rho_2=0$, $\rho_1^\mu=0$ \cite{Denicol:2012cn}.
Therefore, we have $\mathbb{S}^{(0)}_0=\mathbb{S}_2^{(0)}=\{0\}$ and $\mathbb{S}_1^{(0)}=\emptyset$\footnote{Due to the restriction to an uncharged fluid, we do not need to consider the moment $\rho_0^\mu$ related to charge diffusion.}, while $\mathbb{S}_\ell^{(n)}=\emptyset$ for $n>2$. In principle, the transport coefficients in the equations of motion for $\rho_0$ and $\rho^{\mu\nu}_0$ are modified through the coupling to the tensor-polarization moments, known in the nonrelativistic case as the Senftleben effect \cite{hess1971kinetic}. However, it is expected that the modifications of both the conventional transport coefficients and the tensor polarization due to this effect are small \cite{hess1971kinetic}. Furthermore, although the components of the spin tensor should also be treated dynamically \cite{Weickgenannt:2022zxs}, we will not consider them in this work since they do not couple to the tensor-polarization moments.

The tensor polarization in \eq\eqref{tensor_pol0}, when integrated over momentum space, can be expressed in terms of the irreducible moments as
\begin{equation}
\bar{\Theta}^{\mu\nu} \coloneqq \int \d K N(k) \Theta^{\mu\nu}(k) =\frac12 \sqrt{\frac32} \int \d \Sigma_\lambda \left(u^\lambda \psi_1^{\mu\nu}+\psi_0^{\mu\nu,\lambda}\right)\;.\label{tensor_pol}
\end{equation}
In order to keep the degrees of freedom which enter the expression for the tensor polarization \eqref{tensor_pol}, we choose $\mathbb{S}_0^{(2)}=\{1\}$ and $\mathbb{S}_1^{(2)}=\{0\}$ in the moment expansion.

\section{Tensor polarization from shear stress}
\label{sec:tens_pol_from_pi}

Using the truncation procedure outlined in the previous section, the Navier-Stokes limits of \eq\eqref{relax_psi} for $r\in \mathbb{S}_{\ell}^{(2)}$ simply become
\begin{equation}
C_{0}^{\langle\mu\nu\rangle}=0\;, \qquad
C_{-1}^{\langle\mu\nu\rangle,\langle\lambda\rangle}=0\;.\label{eq_psi_mu_nu_lambda}
\end{equation}
When expressing these collision terms through the irreducible moments, we note that, since in this work we only consider parity-conserving interactions, all integrals over $\mathcal{W}$ containing an odd number of spin vectors vanish \cite{Weickgenannt:2022zxs}. This implies that there is no coupling between the moments $\tau_r^{\mu, \mu_1\cdots\mu_\ell}$ and $\psi_r^{\mu\nu,\mu_1\cdots\mu_\ell}$.
The second equation in \eqref{eq_psi_mu_nu_lambda} immediately implies $\psi^{\langle\mu\nu\rangle,\lambda}_0=0$ since there are no tensor structures with the appropriate symmetries. 
On the other hand, the first equation in \eqref{eq_psi_mu_nu_lambda} yields
\begin{equation}
 \sum_{n\in\mathbb{S}^{(2)}_0}\mathcal{C}^{(0)}_{1n}\psi_n^{\langle\mu\nu\rangle} +\sum_{n\in\mathbb{S}^{(0)}_2} \mathcal{D}^{(2)}_{1n} \rho_n^{\mu\nu} =0 \;,\label{NS_psi_0}
\end{equation}
where we linearized the collision term \eqref{C_final}, plugged it into \eqref{irred_coll} and used the expansion for the distribution function \eqref{delta_f_expansion}.
Furthermore, we introduced the collision integrals
\begin{subequations}
\label{eq:coll_ints}
\begin{align}
\mathcal{C}^{(0)}_{1n}&\coloneqq\frac15\int [\d K] f_{0\k} f_{0\k'} \Delta_{\mu\nu,\alpha\beta}\left(\mathcal{M}_{(k\s)(k_1\s_1)}^{\mu\nu,\alpha\beta}\mathcal{H}^{(2,0)}_{k_1 n}+\mathcal{M}_{(k\s)(k_2\s_2)}^{\mu\nu,\alpha\beta}\mathcal{H}^{(2,0)}_{k_2 n}-\mathcal{M}_{(k\s)(k'\s')}^{\mu\nu,\alpha\beta}\mathcal{H}^{(2,0)}_{k' n}-\mathcal{M}_{(k\s)(k\bar{\s})}^{\mu\nu,\alpha\beta}\mathcal{H}^{(2,0)}_{k n}    \right)\;,\\
\mathcal{D}^{(2)}_{1n}&\coloneqq \frac15\int [\d K] f_{0\k} f_{0\k'}  \mathcal{M}_{(k\s)}^{\mu\nu}\left(\mathcal{H}^{(0,2)}_{k_1 n}k_{1,\langle\mu}k_{1,\nu\rangle}+\mathcal{H}^{(0,2)}_{k_2 n}k_{2,\langle\mu}k_{2,\nu\rangle}-\mathcal{H}^{(0,2)}_{k' n}k'_{\langle\mu}k'_{\nu\rangle}-\mathcal{H}^{(0,2)}_{k n}k_{\langle\mu}k_{\nu\rangle}    \right)\;,
\end{align}
\end{subequations}
with $[\d K]\coloneqq \d K_1 \d K_2 \d K'\d K$ and
\begin{subequations}\label{ms}
\begin{align}
\mathcal{M}_{(k\s)}^{\mu\nu}&\coloneqq \frac12 (2\pi\hbar)^4 \delta^{(4)} (k+k'-k_1-k_2)\int \left[\d S\right]\d \bar{S}(k)\, \mathcal{W}K_{\alpha\beta}^{\mu\nu} \s^\alpha \s^\beta \;, \label{m3} \\
\mathcal{M}_{(k_i\s_i)(k_j\s_j)}^{\mu\nu,\alpha\beta}&\coloneqq \frac12 (2\pi\hbar)^4 \delta^{(4)} (k+k'-k_1-k_2) \Xi^{\gamma\delta,\alpha\beta}_{j} \int \left[\d S\right] \d \bar{S}(k)\, \mathcal{W}  K^{\mu\nu}_{i,\rho\sigma} \s_i^\rho \s_i^\sigma \, K^{\zeta\eta}_{j,\gamma\delta} \s_{j,\zeta} \s_{j,\eta}\;.   \label{m4} 
\end{align}
\end{subequations}
Here, we defined $\left[\d S\right]\coloneqq \d S_1(k_1) \d S_2(k_2)  \d S'(k') \d S(k)$, and $K^{\mu\nu}_{i,\alpha\beta}$ denotes the symmetric traceless projector onto the subspace orthogonal to $k_i \in \{k_1,k_2,k',k\}$. Similarly, $\Xi^{\alpha\beta}_{j,\gamma\delta}$ is the tensor introduced in Eq. \eqref{eq:def_Xi} with the momentum $k$ replaced by $k_i$.
A more detailed derivation of Eq. \eqref{NS_psi_0} is provided in Appendix \ref{app:deriv_NS}.

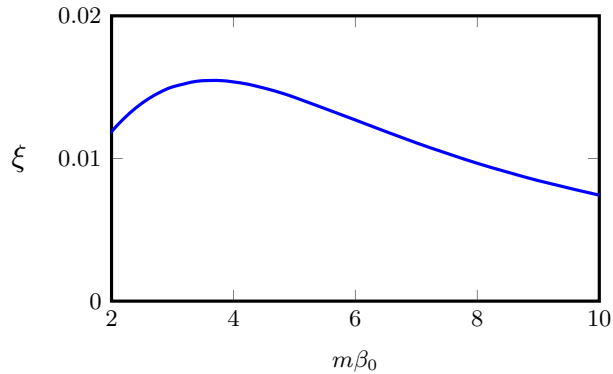
\begin{figure}
    \centering
    \begin{tikzpicture}
    \begin{axis}[enlargelimits=false,width=0.45\textwidth,height=0.3\textwidth,legend style={draw=none},legend pos= north west, ylabel style={rotate=-90}, very thick, yticklabel style = {/pgf/number format/fixed}, scaled y ticks=false, ytick={0,0.01,0.02}, ymin=0, ymax=0.02,xmin=2, xmax=10,
		xlabel=$m\beta_0$,
		ylabel= {\scalebox{1.3}{$\xi$}} ,
		]
		\addplot [blue,smooth] table[x index=0,y index=1,col sep=tab] {coeffdata.txt};
		\end{axis}
    \end{tikzpicture}
    \caption{The coefficient $\xi$ for the case of a four-point interaction, $\mathcal{L}_{\text{int}}\sim (V^{\dagger}\cdot V )^2$.}
    \label{fig:coeff}
\end{figure}

Employing the truncation introduced in Subsec. \ref{subsec:truncation} in \eq\eqref{NS_psi_0} and using that  $\rho_0^{\mu\nu}= \pi^{\mu\nu}$ yields
\begin{equation}
\psi_1^{\langle\mu\nu\rangle}= \xi\,\beta_0 \pi^{\mu\nu}\;,\label{psi_pi}
\end{equation}
where
\begin{equation}
    \xi\coloneqq  -\frac{1}{\beta_0}\frac{\mathcal{D}^{(2)}_{10}}{\mathcal{C}^{(0)}_{11} }
\end{equation}
denotes a coefficient that can only depend on the ratio of mass over temperature $m\beta_0$.
With details relegated to Appendix \ref{app:4pt}, we plot the value of $\xi$ in Fig. \ref{fig:coeff} for the case of a simple four-point interaction. 

Equation \eqref{psi_pi} is one of the main results of this work, showing that the Navier-Stokes values of the moments related to the tensor polarization are determined from collisions. Furthermore, the value of the coefficient $\xi$ is determined solely by local collisions, i.e., the nonlocality of the collision term \eqref{C_final} has no influence on the tensor polarization, provided that the interactions conserve parity.
Note that, neglecting the moments of first order in spin, the deviation of the single-particle distribution function from local equilibrium reads at this point
\begin{equation}
\delta f_{\k\s}=f_{0\k}  \left( -\frac{3}{m^2}\mathcal{H}^{(0,0)}_{\k 0}\Pi +\mathcal{H}^{(0,2)}_{\k 0}k_{\langle\mu}k_{\nu\rangle}\pi^{\mu\nu}+ \xi \beta_0 \mathcal{H}^{(2,0)}_{\k 1} \s_\alpha \s_\beta K^{\alpha\beta}_{\mu\nu} \Xi^{\mu\nu}_{\rho\sigma} \pi^{\rho\sigma} \right)\;.
\label{eq:delta_f_explicit}
\end{equation}

\section{Spin alignment in heavy-ion collisions}
\label{sec:alignment}
We now connect these results to the spin alignment measured in experiments which, in turn, is related to the $00$-element of the spin-density matrix $\rho_{\lambda\lambda'}$ \cite{STAR:2022fan}. In analogy with Ref. \cite{Becattini:2020sww}, one obtains
\begin{equation}\label{density_matrix}
\rho_{\lambda\lambda'}(k)= \frac{\int \d \Sigma_\alpha k^\alpha \epsilon^{(\lambda)\mu}W_{\mu\nu}\epsilon^{*(\lambda')\nu}}{\sum_{\sigma=1}^3 \int \d \Sigma_\alpha k^\alpha \epsilon^{(\sigma)\mu}W_{\mu\nu}\epsilon^{*(\sigma)\nu}}\;.
\end{equation}
The derivation of Eq. \eqref{density_matrix} is provided in Appendix \ref{app:Wigner_to_obs}.
Since we are interested in a diagonal element of the spin-density matrix, with the corresponding polarization vector $\epsilon^{(0)\mu}\coloneqq(0,0,0,1)$ being real, the antisymmetric part of the Wigner function does not contribute. One can verify with the aid of \eq\eqref{tensor_pol0} that the 00-element of \eq\eqref{density_matrix} is given by
\begin{equation}
\rho_{00}(k)=\frac{1}{3}-\sqrt{\frac23} \epsilon^{(0)}_\mu \epsilon^{(0)}_\nu  \Theta^{\mu\nu} (k) \;.
\end{equation}
Using \eqs\eqref{eq:delta_f_explicit}, we arrive at the final expression 
\begin{equation}
\rho_{00}(k)=\frac{1}{3}- \frac{4}{15}\frac{ \int \d \Sigma_\alpha k^\alpha \xi \,\beta_0 f_{0\k}  \mathcal{H}_{\k 1}^{(2,0)} \epsilon^{(0)}_\alpha \epsilon^{(0)}_\beta K_{\mu\nu}^{\alpha\beta} \Xi^{\mu\nu}_{\rho\sigma}\pi^{\rho\sigma} }{\int \d \Sigma_\alpha k^\alpha f_{0\k}\left(1-3\mathcal{H}_{\k 0}^{(0,0)} \Pi/m^2+\mathcal{H}_{\k 0}^{(0,2)}\pi^{\mu\nu}k_{\langle\mu}k_{\nu\rangle}\right)} \;,
\label{eq:final}
\end{equation}
where we used that there is no tensor polarization in local equilibrium which follows from \eq\eqref{f_eq}. The polynomials $\mathcal{H}$ appearing in Eq. \eqref{eq:final} read \cite{Denicol:2012cn}
\begin{equation}
    \mathcal{H}^{(0,0)}_{\k 0} = \frac{1}{I_{00}}\;,\quad \mathcal{H}^{(0,2)}_{\k 0} =\frac{1}{2I_{42}}\;,\quad \mathcal{H}^{(2,0)}_{\k 1} = \frac{15}{8} \frac{I_{00}E_\k -I_{10}}{I_{20}I_{00}-I_{10}^2}  \;.
\end{equation}
Equation \eqref{eq:final} is the main result of our work which shows how vector particles can become tensor polarized due to the presence of shear stress. Since this effect is independent of vorticity, one may choose a quantization axis different from the global angular-momentum direction \cite{ALICE:2019aid,Mohanty:2021vbt,STAR:2022fan}, where the strength might be larger.
It is important to note that the expression \eqref{eq:final} depends on the details of the interaction between particles only through the coefficient $\xi$.

\section{Conclusions}

In this work, starting from quantum kinetic theory and using the method of moments, we have shown that shear stress can induce tensor polarization in an uncharged fluid.
This novel polarization  mechanism is purely related to out-of-equilibrium properties of the system and it is independent of fluid rotation. Thus, one does not need to include nonlocal collisions \cite{Weickgenannt:2020aaf} since such an effect is not determined by the conservation of total angular momentum. 
Our main result is a formula which can be used for quantitative predictions for vector-meson spin alignment in heavy-ion collisions using hydrodynamic simulations.
The present work can be extended by relaxing the assumption of charge-neutrality of the fluid. In fact, particle diffusion will also contribute to the tensor polarization. Furthermore, the method of moments discussed here can be used to derive relativistic dissipative spin-1 hydrodynamics with dynamical spin degrees of freedom.

\section*{Acknowledgments}
When finalizing this work, we became aware of a related study \cite{Li:2022vmb}.
The authors thank E.\ Grossi, U.\ Heinz, X.-G.\ Huang, E.\ Molnár, A.\ Palermo, D.\ H.\ Rischke, A.\ Sadofyev, and Y.\ Yin for enlightening discussions. The work of D.W. and N.W.\ is supported by the
Deutsche Forschungsgemeinschaft (DFG, German Research Foundation)
through the Collaborative Research Center CRC-TR 211 ``Strong-interaction matter
under extreme conditions'' -- project number 315477589 - TRR 211.
D.W.\ and N.W.\ acknowledge the support by the State of Hesse within the Research Cluster ELEMENTS (Project ID 500/10.006).
D.W.\ acknowledges support by the Studienstiftung des deutschen Volkes 
(German Academic Scholarship Foundation) as well as the support through a grant of the
Ministry of Research, Innovation and Digitization, CNCS
- UEFISCDI, project number PN-III-P1-1.1-TE-2021-
1707, within PNCDI III.
N.W. acknowledges the support by the German National Academy of Sciences Leopoldina through the Leopoldina fellowship program with funding code  LPDS 2022-11.

\appendix

\section{Relations between the Wigner function and polarization observables}
\label{app:Wigner_to_obs}
In this Appendix, we prove the relation of the spin-density matrix to the Wigner function, following the same steps as outlined in Ref. \cite{Becattini:2020sww} for spin-1/2 particles. Furthermore, we prove the relation between tensor polarization and the Wigner function reported in the main text.

The spin-density matrix is defined as
\begin{equation}
    \rho_{\lambda\lambda'}(k)\coloneqq\frac{\langle\hat{a}^\dagger_\lambda(k) \hat{a}_{\lambda'}(k)\rangle}{\sum_{\sigma}\langle\hat{a}^\dagger_\sigma(k) \hat{a}_\sigma(k)\rangle}\;.\label{spin_density}
\end{equation}
The goal is to relate the Wigner function 
\begin{equation}
    W^{\mu\nu}(x,k)\coloneqq -\frac{2}{(2\pi\hbar)^4\hbar} \int \d^4 y e^{-ik\cdot y/\hbar} \left\langle :V^{\dagger\mu} (x+y/2) V^\nu (x-y/2) : \right\rangle
\end{equation}
to the averages over creation and annihilation operators appearing in Eq.\ \eqref{spin_density}.
Expressing the fields in terms of creation and annihilation operators 
\begin{equation}
V^\mu (x)\coloneqq \sqrt{\hbar} \sum_\sigma\int \frac{\d^3 k}{(2\pi\hbar)^32k^0} \left[ e^{-\frac{i}{\hbar}k\cdot x } \epsilon^{(\sigma)\mu}(k) \hat{a}_\sigma (k)+e^{\frac{i}{\hbar}k\cdot x } \epsilon^{*(\sigma)\mu}(k) \hat{b}^\dagger_\sigma(k)\right]\label{V_in}
\end{equation}
and inserting them into the Wigner function, we obtain $W^{\mu\nu}=W_+^{\mu\nu}+W_-^{\mu\nu}+W_S^{\mu\nu}$, where $W_{\pm}^{\mu\nu}$ denote the particle and antiparticle contributions, respectively (i.e., their associated momenta are timelike with $k^0>0$ or $k^0<0$), while $W_S^{\mu\nu}$ denotes the Wigner function whose momentum is spacelike. These three quantities read explicitly
\begin{subequations}
\begin{eqnarray}
W_+^{\mu\nu}(x,k)&=&-2\sum_{\sigma,\sigma'} \int \frac{\d^3 p}{(2\pi\hbar)^3 2p^0} \int \frac{\d^3 p'}{(2\pi\hbar)^3 2p'^0} \nonumber\\
&&\times\delta^{(4)}[k-(p+p')/2]e^{i(p-p')\cdot x/\hbar} \epsilon^{*(\sigma)\mu}(p) \epsilon^{(\sigma')\nu}(p') \langle \hat{a}^{\dagger}_\sigma (p) \hat{a}_{\sigma'}(p')  \rangle\label{W_+}\;,\\
W_-^{\mu\nu}(x,k)&=&-2\sum_{\sigma,\sigma'} \int \frac{\d^3 p}{(2\pi\hbar)^3 2p^0} \int \frac{\d^3 p'}{(2\pi\hbar)^3 2p'^0}  \delta^{(4)}[k+(p+p')/2]\nonumber\\
&&\times e^{i(p-p')\cdot x/\hbar} \epsilon^{(\sigma)\mu}(p) \epsilon^{*(\sigma')\nu}(p') \langle \hat{b}^{\dagger}_\sigma (p') \hat{b}^\dagger_{\sigma'}(p)  \rangle\label{W_-}\;,\\
W_S^{\mu\nu}(x,k)&=&-2\sum_{\sigma,\sigma'} \int \frac{\d^3 p}{(2\pi\hbar)^3 2p^0} \int \frac{\d^3 p'}{(2\pi\hbar)^3 2p'^0}  \delta^{(4)}[k-(p-p')/2]
\nonumber\\
&&\times\big[e^{i(p+p')\cdot x/\hbar} \epsilon^{*(\sigma)\mu}(p) \epsilon^{*(\sigma')\nu}(p') \langle \hat{a}^{\dagger}_\sigma (p) \hat{b}^\dagger_{\sigma'}(p')  \rangle\nonumber\\
&&+e^{-i(p+p')\cdot x/\hbar} \epsilon^{(\sigma)\mu}(p') \epsilon^{(\sigma')\nu}(p) \langle \hat{a}_\sigma (p') \hat{b}_{\sigma'}(p)  \rangle\big]\label{W_S}\;,
\end{eqnarray}
\end{subequations}
where we employed that $\langle:\hat{a}_\sigma (p) \hat{a}^\dagger_{\sigma'}(p'):\rangle=\langle\hat{a}^\dagger_{\sigma'}(p')\hat{a}_\sigma (p) \rangle$ due to the bosonic nature of the particles.
Using now that
\begin{equation}
    \int \d \Sigma_\alpha k^\alpha W^{\mu\nu}_+(x,k) \equiv k^0 \int \d^3 x W^{\mu\nu}_+(x,k)=\sum_{\sigma,\sigma'}\delta(k^2-m^2) \Theta(k^0) \epsilon^{*(\sigma)\mu}(p) \epsilon^{(\sigma')\nu}(p') \langle \hat{a}^{\dagger}_\sigma (p) \hat{a}_{\sigma'}(p')  \rangle
\end{equation}
as well as the completeness and orthogonality relations of the polarization vectors,
\begin{equation}
    \epsilon^{*(\lambda)\mu}(k)\epsilon^{(\lambda')}_\mu (k) =-\delta_{\lambda\lambda'}\;, \quad
\sum_{\lambda} \epsilon^{*(\lambda)\mu}(k)\epsilon^{(\lambda)\nu}(k)=-K^{\mu\nu}\;,\label{complete}
\end{equation}
we find the sought-after relation
\begin{equation}
    \langle \hat{a}^\dagger_\lambda(k) \hat{a}_{\lambda'}(k)\rangle= \int \d \Sigma_\alpha k^\alpha \epsilon^{(\lambda)}_\mu(k) W_+^{\mu\nu}(x,k) \epsilon^{*(\lambda')}_\nu(k)\;,
\end{equation}
which lets us express the spin-density matrix of the particles as
\begin{equation}
    \rho_{\lambda\lambda'}(k)=\frac{\int \d \Sigma_\alpha k^\alpha \epsilon^{(\lambda)}_\mu(k) W_+^{\mu\nu}(x,k) \epsilon^{*(\lambda')}_\nu(k)}{\sum_{\sigma} \int \d \Sigma_\alpha k^\alpha \epsilon^{(\sigma)}_\mu(k) W_+^{\mu\nu}(x,k) \epsilon^{*(\sigma)}_\nu(k)}\;.\label{spin_density_wigner}
\end{equation}
Note that a similar relation holds also for the antiparticles.

In the next step we will relate the traceless symmetric components of the Wigner function to the tensor polarization, which is defined as \cite{Leader:2001}
\begin{equation}
    \Theta^{\mu\nu} (k) \coloneqq \frac12 \sqrt{\frac32}\mathrm{Tr}\left[\left( \hat{S}^{(\mu}\hat{S}^{\nu)} +\frac43 K^{\mu\nu}\right)\hat{\rho}(k)\right]\;,\label{tensor_pol_app}
\end{equation}
where $\hat{\rho}(k)$ is the spin-density operator restricted to the four-momentum $k^\mu$, and \begin{equation}
    \hat{S}^\mu \coloneqq -\frac{1}{2m} \epsilon^{\mu\nu\alpha\beta} \hat{J}_{\nu\alpha}\hat{P}_\beta
\end{equation}
denotes the Pauli-Lubanski operator divided by the particle mass \cite{Becattini:2020sww,Speranza:2020ilk}. Here, $\hat{J}^{\mu\nu}$ is the generator of Lorentz transformations, while $\hat{P}^\mu$ generates space-time translations.
From, e.g., Eq.~(14) in Ref.~\cite{Becattini:2020sww} we know that we can represent the matrix elements of the operator $\hat{S}^\mu$ as
\begin{equation}
    \bra{k,\lambda}\hat{S}^\mu\ket{k,\lambda'} =-\frac{1}{2m} \epsilon^{\mu\nu\alpha\beta} k_\nu D^S([k])^{-1} D^S(J_{\alpha\beta}) D^S([k])\;,
\end{equation}
where $D^S(J^{\mu\nu})$ and $D^S([k])$ are the spin-S representation of the total angular-momentum operator and the standard Lorentz boost to the four-momentum $k^\mu$, respectively. 
From this relation we can infer
\begin{equation}
    \Theta^{\mu\nu}(k)=\frac12 \sqrt{\frac32}\left\{\frac12 \epsilon^{\mu\alpha\beta\gamma}\epsilon^{\nu\rho\sigma\lambda} \frac{k_\alpha k_\lambda}{m^2} \mathrm{Tr}\left[D^S([k])^{-1} D^S(J_{\beta\gamma}) D^S(J_{\rho\sigma})D^S([k])\rho(k)\right]+\frac43 K^{\mu\nu}\right\}\;.\label{tensor_pol_2}
\end{equation}
For massive spin-1 particles, we work in the (1/2,1/2) representation of the Lorentz group, where 
\begin{equation}
    D^S(J_{\beta\gamma})^{\mu\nu}=i(g^\mu_\beta g^\nu_\gamma -g^\mu_\gamma g^\nu_\beta)  \;, \quad (D^S(J_{\beta\gamma})D^S(J_{\rho\sigma}))^{\mu\nu}= g^\mu_\beta g^\nu_\rho g_{\gamma\sigma}+g^\mu_\gamma g^\nu_\sigma g_{\beta\rho} -g^\mu_\beta g^\nu_\sigma g_{\gamma\rho}-g^\mu_\gamma g^\nu_\rho g_{\beta\sigma}\;.\label{rep}
\end{equation}
In a basis where the polarization vectors in the particle rest frame [i.e., the frame where $k^{\star\mu}=(m,0,0,0)$] coincide with the cartesian axes, $\epsilon^{(\lambda)\mu}(k^\star)=-g^{\lambda\mu}$, we can express the standard Lorentz transformation as
\begin{equation}
    D^S([k])^{\mu\lambda}=\epsilon^{(\lambda)\mu}(k)\;.
\end{equation}
Inserting this into Eq.\ \eqref{tensor_pol_2} and using the spin-density matrix \eqref{spin_density_wigner} as well as the completeness relation \eqref{complete}, we find
\begin{eqnarray}
\Theta^{\mu\nu}(k)&=&\frac12 \sqrt{\frac32}\left[2 \epsilon^{\mu\alpha\beta\gamma} \epsilon^{\nu\rho\sigma\lambda} \frac{k_\alpha k_\lambda}{m^2} g_{\gamma\sigma}K_{\beta\eta}K_{\rho\zeta} \frac{\int \d \Sigma_{\epsilon} k^\epsilon  W_+^{\eta\zeta}(x,k)}{\int \d \Sigma_\epsilon k^\epsilon K_{\phi \psi} W^{\phi\psi}_+ (x,k)  }+\frac43 K^{\mu\nu}\right]\nonumber\\
&=& \sqrt{\frac32}\left[  (K^\mu_\alpha K^\nu_\beta -K^{\mu\nu}K_{\alpha\beta})\frac{\int \d \Sigma_{\gamma} k^\gamma W_+^{\alpha\beta}(x,k)}{\int \d \Sigma_\gamma k^\gamma K_{\rho\sigma} W^{\rho\sigma}_+ (x,k)  }+\frac23 K^{\mu\nu}\right]\nonumber\\
&=&\sqrt{\frac32}  K^{\mu\nu}_{\alpha\beta}\frac{\int \d \Sigma_{\gamma} k^\gamma W_+^{\alpha\beta}(x,k)}{\int \d \Sigma_\gamma k^\gamma K_{\rho\sigma} W^{\rho\sigma}_+ (x,k)  }\;.
\end{eqnarray}
Translating this expression into integrals over spin space and abbreviating
\begin{equation}
    \int \d \Sigma_\gamma k^\gamma K_{\rho\sigma} W^{\rho\sigma}_+ (x,k)= \int \d \Sigma_\gamma k^\gamma \int \d S(k) f(x,k,\s) \eqqcolon N(k)\;,
\end{equation}
we have
\begin{equation}
    \Theta^{\mu\nu}(k)=\frac12 \sqrt{\frac32} \frac{1}{N(k)} \int \d \Sigma_\gamma k^\gamma \int \d S(k) K^{\mu\nu}_{\alpha\beta} \s^\alpha \s^\beta f(x,k,\s)\;.
\end{equation}

For completeness, we furthermore list the expression for the vector polarization of spin-1 particles, which is defined as
\begin{equation}
    S^\mu (k)\coloneqq \mathrm{Tr}\left[\hat{S}^\mu \hat{\rho}(k)\right]\;.
\end{equation}
Inserting the representation of the total angular momentum operator \eqref{rep}, we obtain
\begin{equation}
    S^\mu (k) =\frac{i}{2} \epsilon^{\mu\nu\alpha\beta} \frac{k_\nu}{m}\frac{\int \d \Sigma_{\gamma} k^\gamma W_{+,\alpha\beta}(x,k)}{\int \d \Sigma_\gamma k^\gamma K_{\rho\sigma} W^{\rho\sigma}_+ (x,k)  }\;,
\end{equation}
which in extended phase space becomes
\begin{equation}
    S^\mu(k) =\frac{1}{N(k)}\int \d \Sigma_\gamma k^\gamma \int \d S(k) \s^\mu f(x,k,\s)\;.
\end{equation}

\section{Derivation of Eq. \eqref{NS_psi_0}}
\label{app:deriv_NS}
Considering the definition of the irreducible moments of the collision integrals \eqref{irred_coll}, Eq. \eqref{eq_psi_mu_nu_lambda} reads explicitly
\begin{align}
    0=C_0^{\langle\mu\nu\rangle}&=  \int \d \Gamma \, K^{\langle\mu\nu\rangle}_{\alpha\beta} \s^\alpha \s^\beta  \mathfrak{C}\nonumber\\
    &= \frac12 \int \left[\d \Gamma\right] \d \bar{S}(k) (2\pi\hbar)^4 \delta^{(4)}(k+k'-k_1-k_2) \mathcal{W}\, K^{\langle\mu\nu\rangle}_{\alpha\beta} \s^\alpha \s^\beta \nonumber\\
    &\times\left[f(x+\Delta_1-\Delta,k_1,\s_1)f(x+\Delta_2-\Delta,k_2,\s_2)-f(x+\Delta'-\Delta,k',\s')f(x,k,\bar{\s})\right] \;,
    \label{eq:C_0_1}
\end{align}
where we abbreviated $\left[\d \Gamma\right]\coloneqq \d \Gamma_1 \d\Gamma_2 \d \Gamma'  \d \Gamma$. Due to our assumption that the interaction conserves parity, all integrals over $\mathcal{W}$ weighted with an odd number of spin vectors vanish \cite{Weickgenannt:2022zxs}, i.e.,
\begin{subequations}
\begin{align}
    \int \left[\d S \right] \d \bar{S}(k) \mathcal{W} \, \s_i^\mu &=0\;,\\
    \int \left[\d S \right] \d \bar{S}(k) \mathcal{W} \, K^{\mu\nu}_{i,\alpha\beta}\s_i^\alpha \s_i^\beta \s_j^\lambda &=0 \;.
\end{align}
\end{subequations}
From these identities we see that only the components of the distribution functions which are proportional to either zero or two spin vectors contribute to Eq. \eqref{eq:C_0_1}. The nonlocal shifts $\Delta_1$, $\Delta_2$, $\Delta'$ and $\Delta$ however are linear in the spin vector $\s^\mu$ \cite{Wagner:2022amr}, which follows from Eqs. \eqref{eq:def_Delta} by considering the symmetries of $M$ together with the assumption that spin effects are at least of order $\mathcal{O}(\hbar)$. 
This implies that neither the nonlocal part of the collision term nor the spin-dependent part of the local-equilibrium distribution function \eqref{f_eq} give a nonvanishing contribution to Eq. \eqref{eq:C_0_1}.
Linearizing the collision term in the deviations from equilibrium, inserting the moment expansion \eqref{delta_f_expansion}, and using the conservation of linear momentum, Eq. \eqref{eq:C_0_1} becomes
\begin{align}
    0&=\frac12 \int \left[\d \Gamma\right] \d \bar{S}(k) (2\pi\hbar)^4 \delta^{(4)}(k+k'-k_1-k_2) \mathcal{W}\, K^{\langle\mu\nu\rangle}_{\alpha\beta} \s^\alpha \s^\beta f_{0\k} f_{0\k'} \nonumber\\
    &\times \sum_{\ell=0}^\infty 
    \Bigg[\sum_{n\in \mathbb{S}_\ell^{(0)}}\left(\mathcal{H}_{\k_1 n}^{(0,\ell)} k_{\langle 1,\mu_1}\cdots k_{1,\mu_\ell\rangle}+\mathcal{H}_{\k_1 n}^{(0,\ell)} k_{\langle 2,\mu_1}\cdots k_{2,\mu_\ell\rangle}-\mathcal{H}_{\k' n}^{(0,\ell)} k'_{\langle\mu_1}\cdots k'_{\mu_\ell\rangle}-\mathcal{H}_{\k n}^{(0,\ell)} k_{\langle\mu_1}\cdots k_{\mu_\ell\rangle}\right)\rho_n^{\mu_1\cdots\mu_\ell}\nonumber\\
    &+\sum_{n\in \mathbb{S}_\ell^{(2)}}\Big(\s_{1,\gamma} \s_{1,\delta} K^{\gamma\delta}_{1,\zeta\eta} \Xi^{\zeta\eta}_{1,\rho\sigma} \mathcal{H}_{\k_1 n}^{(2,\ell)}k_{\langle 1,\mu_1}\cdots k_{1,\mu_\ell\rangle}
    +\s_{2,\gamma} \s_{2,\delta} K^{\gamma\delta}_{2,\zeta\eta} \Xi^{\zeta\eta}_{2,\rho\sigma} \mathcal{H}_{\k_2 n}^{(2,\ell)}k_{\langle 2,\mu_1}\cdots k_{2,\mu_\ell\rangle}\nonumber\\
    &-\s'_\gamma \s'_\delta K'^{\gamma\delta}_{\zeta\eta} \Xi'^{\zeta\eta}_{\rho\sigma} \mathcal{H}_{\k' n}^{(2,\ell)}k'_{\langle\mu_1}\cdots k'_{\mu_\ell\rangle}
    -\bar{\s}_\gamma \bar{\s}_\delta K^{\gamma\delta}_{\zeta\eta} \Xi^{\zeta\eta}_{\rho\sigma} \mathcal{H}_{\k n}^{(2,\ell)}k_{\langle\mu_1}\cdots k_{\mu_\ell\rangle}
    \Big)\psi_n^{\langle\rho\sigma\rangle,\mu_1\cdots\mu_\ell} \Bigg]\nonumber\\
    &\equiv \sum_{\ell=0}^\infty \left[\sum_{n\in \mathbb{S}_\ell^{(0)}} \left(D^{(\ell)}_{1n}\right)^{\mu\nu}_{\mu_1\cdots \mu_\ell} \rho_n^{\mu_1\cdots\mu_\ell}+\sum_{n\in \mathbb{S}_\ell^{(2)}} \left(C^{(\ell)}_{1n}\right)^{\mu\nu}_{\rho\sigma,\mu_1\cdots\mu_\ell} \psi_n^{\langle\rho\sigma\rangle,\langle\mu_1\cdots\mu_\ell\rangle}\right]\;.
    \label{eq:NS_simplified_1}
\end{align}
Here we defined
\begin{subequations}
\begin{align}
    \left(D^{(\ell)}_{1n}\right)^{\mu\nu}_{\mu_1\cdots\mu_\ell} &\coloneqq \frac12 \int \left[\d \Gamma\right] \d \bar{S}(k) (2\pi\hbar)^4 \delta^{(4)}(k+k'-k_1-k_2) \mathcal{W}\, K^{\langle\mu\nu\rangle}_{\alpha\beta} \s^\alpha \s^\beta f_{0\k} f_{0\k'} \nonumber\\
    &\times \left(\mathcal{H}_{\k_1 n}^{(0,\ell)} k_{\langle 1,\mu_1}\cdots k_{1,\mu_\ell\rangle}+\mathcal{H}_{\k_1 n}^{(0,\ell)} k_{\langle 2,\mu_1}\cdots k_{2,\mu_\ell\rangle}-\mathcal{H}_{\k' n}^{(0,\ell)} k'_{\langle\mu_1}\cdots k'_{\mu_\ell\rangle}-\mathcal{H}_{\k n}^{(0,\ell)} k_{\langle\mu_1}\cdots k_{\mu_\ell\rangle}\right)\;,\\
    \left(C^{(\ell)}_{1n}\right)^{\mu\nu}_{\rho\sigma,\mu_1\cdots \mu_\ell} &\coloneqq \frac12 \int \left[\d \Gamma\right] \d \bar{S}(k) (2\pi\hbar)^4 \delta^{(4)}(k+k'-k_1-k_2) \mathcal{W}\, K^{\langle\mu\nu\rangle}_{\alpha\beta} \s^\alpha \s^\beta f_{0\k} f_{0\k'} \nonumber\\
    &\times\Big(\s_{1,\gamma} \s_{1,\delta} K^{\gamma\delta}_{1,\zeta\eta} \Xi^{\zeta\eta}_{1,\rho\sigma} \mathcal{H}_{\k_1 n}^{(2,\ell)}k_{\langle 1,\mu_1}\cdots k_{1,\mu_\ell\rangle}
    +\s_{2,\gamma} \s_{2,\delta} K^{\gamma\delta}_{2,\zeta\eta} \Xi^{\zeta\eta}_{2,\rho\sigma} \mathcal{H}_{\k_2 n}^{(2,\ell)}k_{\langle 2,\mu_1}\cdots k_{2,\mu_\ell\rangle}\nonumber\\
    &-\s'_\gamma \s'_\delta K'^{\gamma\delta}_{\zeta\eta} \Xi'^{\zeta\eta}_{\rho\sigma} \mathcal{H}_{\k' n}^{(2,\ell)}k'_{\langle\mu_1}\cdots k'_{\mu_\ell\rangle}
    -\bar{\s}_\gamma \bar{\s}_\delta K^{\gamma\delta}_{\zeta\eta} \Xi^{\zeta\eta}_{\rho\sigma} \mathcal{H}_{\k n}^{(2,\ell)}k_{\langle\mu_1}\cdots k_{\mu_\ell\rangle}
    \Big) \;.
\end{align}
\end{subequations}
Taking into account that in our truncation $\mathbb{S}_\ell^{(2)}=\emptyset$ for $\ell\geq 2$, it follows from that the tensors defined above must take the following form,
\begin{equation}
    \left(D^{(\ell)}_{1n}\right)^{\mu\nu}_{\mu_1\cdots\mu_\ell}\equiv \mathcal{D}_{1n}^{(2)} \Delta^{\mu\nu}_{\mu_1\mu_2} \delta_{\ell 2}\;, \qquad \left(C^{(\ell)}_{1n}\right)^{\mu\nu}_{\rho\sigma\mu_1\cdots\mu_\ell}\equiv \mathcal{C}_{1n}^{(0)} \Delta^{\mu\nu}_{\rho\sigma} \delta_{\ell 0}\;,
    \label{eq:form_coll_terms}
\end{equation}
where we introduced the scalar coefficients
\begin{equation}
    \mathcal{D}_{1n}^{(2)}\coloneqq \frac15 \Delta^{\mu_1\mu_2}_{\mu\nu}\left(D^{(2)}_{1n}\right)^{\mu\nu}_{\mu_1\mu_2}\;,\qquad \mathcal{C}_{1n}^{(0)}\coloneqq \frac15 \Delta^{\rho\sigma}_{\mu\nu}\left(C^{(0)}_{1n}\right)^{\mu\nu}_{\rho\sigma}\;.
    \label{eq:scalar_coeff_coll}
\end{equation}
The form of the coefficients in Eq. \eqref{eq:form_coll_terms} follows from the fact that the tensors $(D^{(\ell)}_{1n})^{\mu\nu}_{\mu_1\cdots\mu_\ell}$ and $(C^{(\ell)}_{1n})^{\mu\nu}_{\rho\sigma,\mu_1\cdots\mu_\ell}$ have to be orthogonal to $u^\mu$, symmetric and traceless in the indices $(\mu\nu)$, $(\mu_1\cdots\mu_\ell)$, and (in the latter case) $(\rho\sigma)$. The only tensor structures made from $g^{\mu\nu}$ and $u^\mu$ that fulfill these requirements are given by the irreducible projectors of second rank as shown in Eq. \eqref{eq:form_coll_terms}.
Inserting Eqs. \eqref{eq:form_coll_terms} and \eqref{eq:scalar_coeff_coll} into Eq. \eqref{eq:NS_simplified_1}, we arrive at Eq. \eqref{NS_psi_0} in the main text.

\section{Calculations for a four-point interaction}
\label{app:4pt}
Considering a simple four-point interaction characterized by a dimensionless coupling strength $G$, 
\begin{equation}
    \mathcal{L}_{\text{int}} \coloneqq  \hbar G  (V^{\dagger}\cdot V)^2\;,
\end{equation}
we compute the transfer-matrix elements at leading order \cite{DeGroot:1980dk, Wagner:2022amr}
\begin{align}
\bra{k,k';\lambda,\lambda'} \hat{t} \ket{k_1,k_2;\lambda_1,\lambda_2} &= \frac{1}{\hbar}\bra{k,k';\lambda,\lambda'} :\mathcal{L}_{\text{int}} (0):\ket{k_1,k_2;\lambda_1,\lambda_2}\nonumber\\
&= 2 \hbar^2 G \Big\{\left[\epsilon^{*(\lambda')}_\alpha(k') \epsilon^{(\lambda_1)\alpha} (k_1)\right]\left[\epsilon^{*(\lambda)}_\beta (k) \epsilon^{(\lambda_2)\beta} (k_2)\right]\nonumber\\
&\qquad \quad\; +\left[\epsilon^{*(\lambda')}_\alpha (k') \epsilon^{(\lambda_2)\alpha} (k_2) \right]\left[\epsilon^{*(\lambda)}_\beta (k) \epsilon^{(\lambda_1)\beta} (k_1)\right]\Big\}\;,
\end{align}
where we used the free-field representation of the vector fields
\begin{equation}
    V^\mu (0) =\sqrt{\hbar}\sum_{\sigma'} \int \frac{\d^3 \mathbf{k}'}{(2\pi\hbar)^3 2k'^0} \hat{a} (k',\sigma')\epsilon^{(\sigma')\mu}(k')\;.
\end{equation}
Recalling the relationship \eqref{eq:rel_t_M} between the vertices $M$ and the transfer-matrix elements, we find
\begin{equation}
    M^{\mu\nu\alpha\beta}= 2 \hbar^2 G\left(g^{\mu\alpha}g^{\nu\beta} +g^{\mu\beta}g^{\nu\alpha}\right)\;.
    \label{eq:M_4pt}
\end{equation}
Using the identities
\begin{subequations}
\begin{align}
    \int \d S (k) h^{\mu\nu} (k,\s)&=K^{\mu\nu} \;,\\
    \int \d S (k) H^{\mu\nu} (k,\s) &=K^{\mu\nu}\;,\\
    \int \d S (k) K_{\alpha\beta}^{\rho\sigma} \s_\rho \s_\sigma  h^{\mu\nu} (k,\s) &=  \frac85 K^{\mu\nu}_{\alpha\beta} \;,\\
    \int \d S (k) K_{\alpha\beta}^{\rho\sigma} \s_\rho \s_\sigma  H^{\mu\nu} (k,\s) &=  K^{\mu\nu}_{\alpha\beta}\;,
\end{align}
\end{subequations}
we are able to perform the integrals over spin space in Eqs. \eqref{ms}, obtaining
\begin{subequations}
\label{eq:ms_spin_space_ints}
\begin{align}
    \int [\d S] \d \bar{S} (k) \mathcal{W} K^{\mu\nu}_{\alpha\beta} \s^\alpha \s^\beta &= \frac{(2\pi\hbar)^3}{16} M^{\gamma_1\gamma_2\delta_1\delta_2}M^{\zeta_1\zeta_2\eta_1\eta_2} K_{1,\gamma_1\eta_1} K_{2,\gamma_2\eta_2} K'_{\zeta_2\delta_2} K_{\zeta_1\delta_1}^{\mu\nu} \;,\\
    \int [\d S] \d \bar{S} (k) \mathcal{W} K^{\mu\nu}_{\rho\sigma} \s^\rho \s^\sigma K^{\gamma\delta}_{1,\zeta\eta} \s_1^\zeta \s_1^\eta &= \frac85 \frac{(2\pi\hbar)^3}{16} M^{\gamma_1\gamma_2\delta_1\delta_2}M^{\zeta_1\zeta_2\eta_1\eta_2} K^{\gamma\delta}_{1,\gamma_1\eta_1} K_{2,\gamma_2\eta_2} K'_{\zeta_2\delta_2} K_{\zeta_1\delta_1}^{\mu\nu} \;,\\
     \int [\d S] \d \bar{S} (k) \mathcal{W} K^{\mu\nu}_{\rho\sigma} \s^\rho \s^\sigma K^{\gamma\delta}_{2,\zeta\eta} \s_2^\zeta \s_2^\eta &= \frac85\frac{(2\pi\hbar)^3}{16} M^{\gamma_1\gamma_2\delta_1\delta_2}M^{\zeta_1\zeta_2\eta_1\eta_2} K_{1,\gamma_1\eta_1} K^{\gamma\delta}_{2,\gamma_2\eta_2} K'_{\zeta_2\delta_2} K_{\zeta_1\delta_1}^{\mu\nu} \;,\\
      \int [\d S] \d \bar{S} (k) \mathcal{W} K^{\mu\nu}_{\rho\sigma} \s^\rho \s^\sigma K'^{\gamma\delta}_{\zeta\eta} \s'^\zeta \s'^\eta &= \frac85\frac{(2\pi\hbar)^3}{16} M^{\gamma_1\gamma_2\delta_1\delta_2}M^{\zeta_1\zeta_2\eta_1\eta_2} K_{1,\gamma_1\eta_1} K_{2,\gamma_2\eta_2} K'^{\gamma\delta}_{\zeta_2\delta_2} K_{\zeta_1\delta_1}^{\mu\nu} \;,\\
       \int [\d S] \d \bar{S} (k) \mathcal{W} K^{\mu\nu}_{\rho\sigma} \s^\rho \s^\sigma K^{\gamma\delta}_{\zeta\eta} \bar{\s}^\zeta \bar{\s}^\eta &= \frac85\frac{(2\pi\hbar)^3}{16} M^{\gamma_1\gamma_2\delta_1\delta_2}M^{\zeta_1\zeta_2\eta_1\eta_2} K_{1,\gamma_1\eta_1} K_{2,\gamma_2\eta_2} K'_{\zeta_2\delta_2} K_{\zeta_1\rho}^{\mu\nu} g^{\rho\sigma} K^{\gamma\delta}_{\sigma\delta_1} \;.
\end{align}
\end{subequations}
Inserting the vertices given in Eq. \eqref{eq:M_4pt} into Eqs. \eqref{eq:ms_spin_space_ints}, we perform the remaining momentum integrals \eqref{eq:coll_ints} via slightly modifying a method outlined in Chapter XIII of Ref. \cite{DeGroot:1980dk}, which we now briefly outline. 

The basic idea consists in separating the integrals in Eqs. \eqref{eq:coll_ints} into a sum of elementary collision integrals
\begin{equation}
    J^{(a,b,d,e,f)}\coloneqq \int [\d K] e^{-\beta P_T\cdot u} (P_T^2)^a (P_T\cdot u)^b (Q\cdot u)^d (Q'\cdot u)^e (-Q\cdot Q')^f \delta^{(4)}(k+k'-k_1-k_2)\;,
    \label{eq:basic_coll}
\end{equation}
where the momenta $k,k',k_1$, and $k_2$ can be expressed in terms of the total momentum $P_T$ and the relative momenta $Q$, $Q'$ via
\begin{subequations}
\begin{align}
k^\mu &= \frac12 \left(P_T^\mu+Q^\mu\right) \;,\\
k'^\mu &= \frac12 \left(P_T^\mu-Q^\mu\right) \;,\\
k_1^\mu &= \frac12 \left(P_T^\mu+Q'^\mu\right) \;,\\
k_2^\mu &= \frac12 \left(P_T^\mu-Q'^\mu\right) \;.
\end{align}
\end{subequations}
Next we follow the steps in Ref. \cite{DeGroot:1980dk} and make use of the integral
\begin{equation}
    \int_z^\infty \d y\, \left(y^2-z^2\right)^{b-1/2} y^{a} e^{-y} = z^{a+2b}\sum_{j=0}^{b} (-1)^j \binom{b}{j}  \mathrm{Ki}_{2j-2b-a}(z)\;,
    \label{eq:int_Ki}
\end{equation}
where $\mathrm{Ki}_r(z)$ denotes the Bickley-Naylor function of order $r$ \cite{Abramowitz66}. The result for the basic integral \eqref{eq:basic_coll} then reads
\begin{align}
    J^{(a,b,d,e,f)}&=\beta^{-4-2a-b-d-e-2f} \frac{16\pi^3}{(2\pi\hbar)^{12}} \sum_{g=0}^{\min (d,e)} K(d,e,g) \sigma^{(f,g)}   \sum_{h=0}^{\frac{d+e}{2}+1} \binom{\frac{d+e}{2}+1}{h} (-1)^h \nonumber\\
    &\times \int_{2z}^\infty \d v \left[v^2-(2z)^2\right]^{(d+e)/2+f+1} v^{2(a-1)+b+3} \mathrm{Ki}_{-b-d-e-2+2h}(v)\;,
    \label{eq:basic_coll_solution}
\end{align}
where we introduced the following factors,
\begin{subequations}
\begin{align}
    K(d,e,g)&\coloneqq \begin{dcases} \frac{d! e!}{(d-g)!!(d+g+1)!!(e-g)!!(e+g+1)!!} \;, \quad &\text{if} \; (d-g), \,(e-g) \; \text{even}\;,\\
    0\;, \quad &\text{otherwise}\;,
    \end{dcases}\\
    \sigma^{(f,g)} &\coloneqq \begin{dcases}
        (2g+1)\frac{f!\, 2^g}{(f+g+1)!} \frac{\left(\frac{f+g}{2}\right)!}{\left(\frac{f-g}{2}\right)!}\;, \quad &\text{if} \; (f-g) \; \text{even}\;,\\
        0\;, \quad &\text{otherwise}\;.
    \end{dcases}
\end{align}
\end{subequations}
The remaining task then consists in expanding the integrals \eqref{eq:coll_ints} as sums of the basic integrals \eqref{eq:basic_coll_solution}. 
Note that the tensors $\Xi^{\mu\nu}$, $\Xi^{\mu\nu}_{\alpha\beta}$ do not allow for a straightforward expression in terms of polynomials of $P_T$, $Q$ and $Q'$. This is the case because of the factors of energy appearing in the denominator, leading to
\begin{equation}
    \Xi^{\mu\nu}=\Delta^{\mu\nu}+\frac{(P_T^{\langle\mu\rangle}+Q^{\langle\mu\rangle})(P_T^{\langle\nu\rangle}+Q^{\langle\nu\rangle})}{(P_T\cdot u+ Q\cdot u)^2}\;,
\end{equation}
and similarly for $\Xi^{\mu\nu}_{\alpha\beta}$. In order to bring these terms into the form required by Eq. \eqref{eq:basic_coll} as well, we expand them around the nonrelativistic limit (formally equivalent to taking the limit $k^\mu\simeq (m,\mathbf{0})^\mu$), leading to
\begin{equation}
    \Xi^{\mu\nu}\simeq \Delta^{\mu\nu}\;,\quad \Xi^{\mu\nu}_{\alpha\beta} \simeq \Delta^{\mu\nu}_{\alpha\beta}\;.
\end{equation}
The plot \ref{fig:coeff} is generated with this leading-order approximation, which our tests suggest is reasonable for the covered values of $z$, with accuracy increasing towards larger values of $z$.

\bibliography{biblio_paper_long.bib}

\end{document}